\documentclass[onecolumn,10pt]{IEEEtran}
\usepackage{amsfonts,amsmath,amssymb,amsthm,caption,algorithmic}
\usepackage{amsbsy}
\usepackage{graphicx,multirow,bm}
\usepackage{color}
\usepackage{booktabs}
\usepackage[ruled]{algorithm2e}
\makeatletter
\newtheoremstyle{mythm}{3pt}{3pt}{}{16pt}{\bfseries}{:}{.5em}{}
\theoremstyle{mythm}
\newtheorem{theorem}{Theorem}
\begin{document}
\title{Freshness-Optimal Caching for Information Updating Systems with Limited Cache Storage Capacity
\author{Haibin Xie, Minquan Cheng, Yongbing Zhang}
\thanks{H. Xie is with School of Science, Guilin University of Aerospace Technology, Guilin 541004, China (e-mail:xiehaibin426@163.com)}
\thanks{M. Cheng is with Guangxi Key Lab of Multi-source Information Mining $\&$ Security, Guangxi Normal University,
Guilin 541004, China (e-mail: $\{$chengqinshi,jjiang2008$\}$@hotmail.com).}
\thanks{Y. Zhang is Graduate School of Systems and Information Engineering University of Tsukuba, Japan (email: ybzhang@sk.tsukuba.ac.jp)}
}
\date{}
\maketitle

\begin{abstract}  
In this paper, we investigate a cache updating system with a server containing $N$ files, $K$ relays and $M$ users. The server keeps the freshest versions of the files which are updated with fixed rates. Each relay can download the fresh files from the server in a certain period of time. Each user can get the fresh files from any relay as long as the relay has stored the fresh versions of the requested files. Due to the limited storage capacity and updating capacity of each relay, different cache designs will lead to different average freshness of all updating files at users. In order to keep the average freshness as large as possible in the cache updating system, we formulate an average freshness-optimal cache updating problem (AFOCUP) to obtain an optimal cache scheme. However, because of the nonlinearity of the AFOCUP, it is difficult to seek out the optimal cache scheme. As a result, an linear approximate model is suggested by distributing the total update rates completely in accordance with the number of files in the relay in advance. Then we utilize the greedy algorithm to search the optimal cache scheme that is satisfied with the limited storage capacity of each relay. Finally, some numerical examples are provided to illustrate the performance of the approximate solution.
\end{abstract}

\section{INTRODUCTION}
With emergence of new services and application scenarios, such as YouTube and Netflix, massive device connectivity is becoming a main challenge for existing Internet traffic. To alleviate the latency of data transmission between the servers on the Internet and end users, caching is a promising technology by storing the required files or contents at {\it relay servers} (or simply as {\it relays} hereafter) nearby the users so that the users can retrieve the cached data with low latency \cite{SGDMC}-\cite{DAFY2}. Moreover, with the popularity of technologies such as autonomous driving, augmented reality, social networking, high-frequency automated trading, and online gaming, time sensitive information has become much more important than ever. Age of information has been proposed as an effective metric to quantify the freshness of information in communication networks. There have been lots of efforts on Age of information such as social networks \cite{ICM}, web crawling \cite{CG}-\cite{KPLH}, queueing networks \cite{BSS}-\cite{SU}, caching systems \cite{GCSI}-\cite{PAGBAR},  scheduling in networks \cite{KSUSM}-\cite{BU4}, multi-hop multicast networks \cite{ZSY}-\cite{BSU3}, reinforcement learning \cite{ZCV}-\cite{CGG} and so on.

The freshness of cached data is often used as a metric to evaluate the update effect of contents or files requested by users in a cache updating system. At present, there are some preliminary studies on freshness such as \cite{KPLH}, \cite{YCYW}, \cite{BU1}-\cite{BU2}. In \cite{KPLH}, Kolobov et al. studied the problem of finding optimal crawl rates to keep the information in a search engine fresh while maintaining the constraints on crawling rates imposed by the web sites and also the total crawl rate constraint of the search engine. Yates et al. in \cite{YCYW} considered an age metric to quantify the freshness and then proposed a model where a resource constrained remote server wants to keep the requested contents at a local cache as fresh as possible. They showed that the update rates of the files should be chosen proportional to the square roots of their popularity indices. Different from reference \cite{YCYW} where the freshness of the local cache is considered, Bastopcu and Ulukus in \cite{BU1} considered the freshness at the user which is connected to the source via a single cache or multiple caches. Furthermore, the freshness in \cite{BU1} is different from the traditional age metric used in \cite{YCYW}. However, the caching capacity of the cache in the system model considered by \cite{BU1} is unlimited, which may not be possible in practice. Therefore, Bastopcu and Ulukus in \cite{BU2} made further efforts to consider a cache updating system that consists of a source, a cache with limited caching capacity, and discussed the trade-off between the storing the files at the cache and obtaining the files directly from the source.

Usually, a number of relays are deployed in the edge networks and the files required by a user can be cached at one or multiple nearby relays. However, besides the caching capacity, the update rate of cached data at a relay is limited by the network bandwidth in practice. Servers on the Internet keep the fresh versions of all the files, which are updated with known rates. On the other hand, the files  downloaded by and stored at a relay are valid (fresh) only within a certain period of times. Users retrieve the files from nearby relays that may be valid or outdated. This implies that a user may get a fresh or an outdated file depending on the file freshness. In this paper, we prefer to design a cache updating scheme that allocates file caches at the relays nearby the users so as to maximize the average freshness of files accessed by users.

We first provided an average freshness-optimal caching updating problem (AFOCUP). Through analyzing the AFOCUP problem, the optimal file cache scheme, which produced the maximal average freshness of all files at users in the cache updating system, was obtained. The uncertainty of update rate of each file at relays leads to the nonlinearity of the AFOCUP. By distributing the total update rate completely in accordance with the number of files in each relay, we linearize the original objective function and then get an approximation of the AFOCUP. With the help of the Greedy Algorithm, an optimal cache scheme for the approximation of AFOCUP can be obtained. Moreover, we analyse the relationship between the objective function and two different update rates of files at users and server and find the optimal average freshness is monotonically decreasing on file update rate at the server as well as increasing at users.

\section{NETWORK SYSTEM MODEL}
\subsection{System Scenario}
\label{subsect-System}
\begin{figure}[h]
\centering
\includegraphics[width=2.4in]{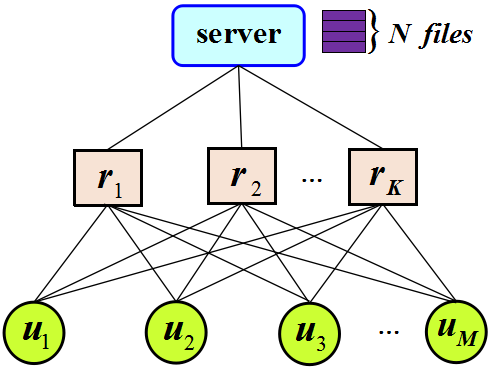}
\vskip 0.2cm
\caption{Cache updating system }\label{figure1}
\end{figure}

In this paper, we consider a cache updating system that consists of a server, multiple relays and users. That is, there is a server $r_0$, $K$ relays $\mathcal{R}=\{r_{1},r_{2},\ldots,r_{K}\}$, and $M$ users $\mathcal{U}=\{u_1,u_2,\ldots,u_M\}$, as shown in Fig. \ref{figure1}. We assume that different users store different files and each file is only cached by one of these relays, and both the storage capacity and total update rate are limited for each relay in the updating system.

It is assumed that $r_{0}$ stores $N$ files denoted by $\mathcal{F}=\{f_{1},f_{2},\ldots,f_{N}\}$, each relay can cache at most $C_{k}$ files, and each user $u_{i}$ can cache at most $n_{i}$ files, i.e., $\mathcal{F}_{u_{i}}=\{f_{u_{i},1},f_{u_{i},2},\ldots,f_{u_{i},n_{i}}\}\subseteq F$. Clearly $C_{k}, n_i\leq N$ always holds for all $k$ and $i$. The files update in the server, relay and user's cache are satisfied the following assumptions.
\begin{itemize}
\item The server keeps the freshest version of $N$ files where $f_{j}$ is updated with exponential inter-arrival times with rates $\lambda_{r_{0},f_{j}}$. The files update at the server are independent of each other;
\item When a file is updated at the server, the stored version of the same file at the relay become outdated. When the relay gets an update for an outdated file, the updated file in the relay becomes fresh again until the next update arrival at the server. Each relay $r_{k}$ can download the latest versions of stored files from the server with delivery cost. For different files, the delivery costs are usually different on account of delay period. Therefore, the inter-update request times of relay $r_{k}$ for file $f_{j}$ are exponential with rate $\lambda_{r_{k},f_{j}}$ in \cite{BU1}. Due to the resource-constrained, each relay $r_{k}$ is subject to a total update rate constraint $G_{k}$, i.e., $\sum_{f_{j}\in F} \lambda_{r_{k},f_{j}}\leq G_{k}$.
\item When a file is updated at the server or relay, the stored version of the same file at the user become outdated. Each user can obtain the latest version from the relay which contains its requested update file. Similar delivery cost consideration for each user, the inter-update request times of user $u_{i}$ for file $f_{j}$ are exponential with rate $\lambda_{u_{i},f_{j}}$. Moreover, each user only choose one of the relays to download the fresh file.
\end{itemize}

It is noted that when the user requests an update for an outdated file, it might still receive an outdated version if the file at the relay is not fresh. Since the relays and the users are unaware of the file updates at the server, they do not know whether they have the freshest versions of the files or not. Thus, they may still request an update even though they have the freshest version of a file. In order to keep track of the freshness at the users, Bastopcu and Ulukus in \cite{BU1} provide an analytical expression for the freshness of the files at the user, and then utilized the freshness index to evaluate the updating effect of the file at users.

For convenience, we denote $F_{r_{k},u_{i},f_{j}}$ as the freshness of file $f_{j}$ at the user $u_{i}$ when the file $f_{j}$ is updated successfully at the relay $r_{k}$. The freshness $F_{r_{k},u_{i},f_{j}}$ is related to the above three update rates $\lambda_{r_{0},f_{j}}$, $\lambda_{r_{k},f_{j}}$ and $\lambda_{u_{i},f_{j}}$. The details of the freshness will be illustrated in the following section. According to \cite{BU1}, for file $f_{j}$ at the user $u_{i}$, different choices of the relays will produce different freshness $F_{r_{k},u_{i},f_{j}}$. With respect to the above cache updating system in Fig.1, we will provide a cache updating strategy to illustrate the assignment of these relays to cache the files at the users in the following section.

\subsection{Cache Updating Strategy}
For file $f_{j}$ at the user $u_{i}$, if it is cached by relay $r_{k}$, it will be endowed with a appropriate update rate $\lambda_{r_{k},f_{j}}$. As a result, the freshness $F_{r_{k},u_{i},f_{j}}$ will change along with the changes of the update rate at different relays. The caching updating strategy of the caching updating system introduced in Subsection \ref{subsect-System} can be informed as follows.

Let $X$ be a cache updating scheme, which is a binary variable set denoted by\\
\[X=\left\{x_{r_{k},u_{i},f_{j}}: r_{k}\in\mathcal{R};~u_{i}\in\mathcal{U};~f_{j}\in\mathcal{F}_{u_{i}}\right\},\]
where $x_{r_{k},u_{i},f_{j}}$ indicates the caching status of file $f_{j}$ in relay $r_{k}$, i.e., $x_{r_{k},u_{i},f_{j}}=1$ if it is cached by relay $r_{k}$ and user $u_{i}$ can download the latest version from the relay, otherwise $x_{r_{k},u_{i},f_{j}}=0$ if it is not cached in relay $r_{k}$ and certainly user $u_{i}$ can not get the updated file from relay $r_{k}$.

We assume that each file $f_{j}\in\mathcal{F}_{u_{i}}$ can be updated by only one of these relays successfully. Then the following equivalent constraints are satisfied
\begin{eqnarray*}
\begin{split}
&~\sum_{k=1}^{K}x_{r_{k},u_{i},f_{j}}=1,~~~ u_{i}\in \mathcal{U},~f_{j}\in\mathcal{F}_{u_{i}},\\
&\sum_{f_{j}\in\mathcal{F}_{u_{i}}}\sum_{k=1}^{K}x_{r_{k},u_{i},f_{j}}=n_{i},~~~ u_{i}\in\mathcal{U}.~~~~~~~
\end{split}
\end{eqnarray*}Due to the limited cache storage capacity, each relay is subject to the constraint condition, i.e.
\begin{eqnarray*}
~~~\sum_{i=1}^{M}\sum_{f_{j}\in\mathcal{F}_{u_{i}}}x_{r_{k},u_{i},f_{j}}\leq C_{k},~~~ k=1,2,\ldots,K,
\end{eqnarray*}Moreover, each relay $r_{k}$ is satisfied with the total update rate constraint $G_{k}$, i.e.
\begin{eqnarray*}
~~~~~~~~~~~~~\sum_{i=1}^{M}\sum_{f_{j}\in\mathcal{F}_{u_{i}}}\lambda_{r_{k},u_{i},f_{j}}\cdot x_{r_{k},u_{i},f_{j}}\leq G_{k},~~~ k=1,2,\ldots,K.
\end{eqnarray*}In general, different users have different files as well as different preferences for the same file. For simplicity, it is assumed that the requests for all files in the server follow the same distribution. Let file $f_{j}$ be requested with probability $p_{f_{j}}$. Clearly $\sum_{f_{j}\in\mathcal{F}~}$$p_{f_{j}}=1$. For user $u_{i}$, the probability of file $f_{j}\in\mathcal{F}_{u_{i}}$ updated is defined as
\[p_{u_{i},f_{j}}=\frac{p_{f_{j}}}{\sum_{f_{j}\in\mathcal{F}_{u_{i}}}p_{f_{j}}}.\]
In addition, the user's decision on relay selection is affected by some actual factors such as transmission times, storage capacity of the relay or user, total update rates of the relay etc. So each user usually has different preferences for different relays. The probability, of which $u_{i}$ requests the updating file from the relay $r_{k}$, is denoted by $p_{r_{k},u_{i}}$. Then we have
\begin{eqnarray*}
\sum_{k=1}^{K} p_{r_{k},u_{i}}=1, ~~u_{i}\in\mathcal{U}.
\end{eqnarray*}It is noted that freshness is an important metric to quantify the performance of the cache updating system. The detail of freshness is described in Section III. Next, for the cache updating system described as Fig.\ref{figure1}, we will propose an analytic expression on the average freshness of all files for all users. For file $f_{j}\in\mathcal{F}_{u_{i}}$, the expected freshness $F_{u_{i},f_{j}}$ for all relays is represented by
\begin{eqnarray*}
F_{u_{i},f_{j}}=\sum_{k=1}^{K}p_{r_{k},u_{i}}\cdot F_{r_{k},u_{i},f_{j}}\cdot x_{r_{k},u_{i},f_{j}}.
\end{eqnarray*} Then, for all files at the user $u_{i}$, we can get the expected freshness $F_{u_{i}}$ as follows
\begin{eqnarray*}
\begin{split}
F_{u_{i}}&=\sum_{f_{j}\in\mathcal{F}_{u_{i}}}p_{u_{i},f_{j}}\cdot F_{u_{i},f_{j}}\\
&=\sum_{f_{j}\in\mathcal{F}_{u_{i}}}p_{u_{i},f_{j}}\sum_{k=1}^{K}p_{r_{k},u_{i}}\cdot F_{r_{k},u_{i},f_{j}}\cdot x_{r_{k},u_{i},f_{j}}.
\end{split}
\end{eqnarray*}where $p_{u_{i},f_{j}}$ is the probability of file $f_{j}$ requested by user $u_{i}$.

Based on above description and discussion, for all users in the cache updating system, the average freshness of all files is expressed as
\begin{eqnarray}
~~~~~F&=&\frac{1}{M}\sum_{i=1}^{M}F_{u_{i}} \notag  \\
~~~~~&=&\frac{1}{M}\sum_{i=1}^{M}\sum_{f_{j}\in\mathcal{F}_{u_{i}}}p_{u_{i},f_{j}}\sum_{k=1}^{K}p_{r_{k},u_{i}}\cdot F_{r_{k},u_{i},f_{j}}\cdot x_{r_{k},u_{i},f_{j}}. \label{eq-maxim-1}
\end{eqnarray}

\section{PROBLEM FORMULATION AND COMPLEXITY ANALYSIS}

\subsection{Problem Formulation}
In the cache updating system, both the update rate $\lambda_{r_{0},f_{j}}$ at the server $r_{0}$ and $\lambda_{u_{i},f_{j}}$ at the user $u_{i}$ usually remain unchanged during a certain period. However, the update rate $\lambda_{r_{k},f_{j}}$ at the relay $r_{k}$ is uncertain as a result of diversity of cache schemes. With regarding to the total update rates $G_{k}$, each relay needs to allocate $\lambda_{r_{k},f_{j}}$ in accordance with the number of its stored files in specific cache status. That is to say, the update rate $\lambda_{r_{k},f_{j}}$ will be endowed with different values in different cache schemes to make full use of total update rates $G_{k}$. In this section, we investigate the optimal caching update strategy to maximize the average freshness $F$ over all files at the users under some necessary constraints. Thus, the maximum average freshness-optimal caching updating problem (AFOCUP) can be formulated as
\begin{eqnarray}
&\max &F=\frac{1}{M}\sum_{i=1}^{M}\sum_{f_{j}\in\mathcal{F}_{u_{i}}}p_{u_{i},f_{j}}\sum_{k=1}^{K}p_{r_{k},u_{i}}\cdot F_{r_{k},u_{i},f_{j}}\cdot x_{r_{k},u_{i},f_{j}}, \notag \\
&s.t.& \sum_{i=1}^{M}\sum_{f_{j}\in\mathcal{F}_{u_{i}}}\lambda_{r_{k},f_{j}}\cdot x_{r_{k},u_{i},f_{j}}\leq G_{k},~~k=1,2,\ldots,K,  \label{eq-maxim-2}\\
 & & \sum_{i=1}^{M}\sum_{f_{j}\in\mathcal{F}_{u_{i}}}x_{r_{k},u_{i},f_{j}}\leq C_{k},~~~~~~~~~~~k=1,2,\ldots,K,\label{eq-maxim-3}\\
 & & \sum_{k=1}^{K}x_{r_{k},u_{i},f_{j}}=1, ~~~~~~~~~~~~~~~~~~~~~u_{i}\in\mathcal{U},~f_{j}\in\mathcal{F}_{u_{i}},\label{eq-maxim-4} \\
 & & \sum_{f_{j}\in\mathcal{F}_{u_{i}}}\sum_{k=1}^{K}x_{r_{k},u_{i},f_{j}}=n_{i},~~~~~~~~~~~~u_{i}\in\mathcal{U},\label{eq-maxim-5}\\
 & & x_{r_{k},u_{i},f_{j}}\in X. \label{eq-maxim-6}
\end{eqnarray}

In order to solve the AFOCUP problem, we need to illustrate the analytical expression for the freshness of the files at the user in the above optimal model.

\subsection{Freshness Index}
As one of the most important indexes, the freshness is used to evaluate the performance of the cache updating system in reference\cite{BU1,BU2}. In this section, we will provide the analytical expression for the freshness of the files at the users. Bastopcu and Ulukus in\cite{BU1} defined the $freshness~function$~$f_{u}(f_{j},t)$ at the user $u$ as follows
\begin{eqnarray*}
f_{u}(f_{j},t)=\left\{
\begin{array}{cc}
1,&\mbox{if $f_{j}$ is fresh at time $t$,}\\
0,&\mbox{otherwise.~~~~~~~~~~~~~~~~}
\end{array}
\right.
\end{eqnarray*}
\noindent
i.e., the instantaneous freshness function is a binary function taking values of fresh, ¡°1¡±, or not fresh ¡°0¡±, at any time $t$. We can define the average freshness of file $f_{j}$ at the user $u$ as follows.
\[
F_{u}(f_{j})=\lim\limits_{T\to\infty}\frac{1}{T}\int_{0}^{T}f_{u}(f_{j},t)dt.
\]
Denote the time interval between the $k$th and the $(k+1)$th successful updates for file $f_{j}$ at the user as the $k$th update cycle by $I(f_{j},k)$. We denote the time duration when file $f_{j}$ at the user is fresh during the $k$th update cycle as $T(f_{j},k)$. Then, the average freshness $F_{u}(f_{j})$ can be written as
\[
F_{u}(f_{j})=\lim\limits_{T\to\infty}\frac{N}{T}\left[\frac{1}{N}\sum \limits_{k=1}^{N}T(f_{j},k)\right]=\frac{E[T(f_{j})]}{E[I(f_{j})]},
\]
where $N$ is the number of update cycles in the time duration $T$. When file $f_{j}\in\mathcal{F}_{u_{i}}$ is updated by relay $r_{k}$, the expression of $freshness$ is formulated as (see\cite{BU1})
\begin{eqnarray}
F_{r_{k},u_{i},f_{j}}=\frac{\lambda_{u_{i},f_{j}}}{\lambda_{u_{i},f_{j}}+\lambda_{r_{0},f_{j}}}\cdot \frac{\lambda_{r_{k},f_{j}}}{\lambda_{r_{k},f_{j}}+\lambda_{r_{0},f_{j}}}. \label{eq-maxim-7}
\end{eqnarray}

\begin{figure}[h]
\centering
\includegraphics[width=3in]{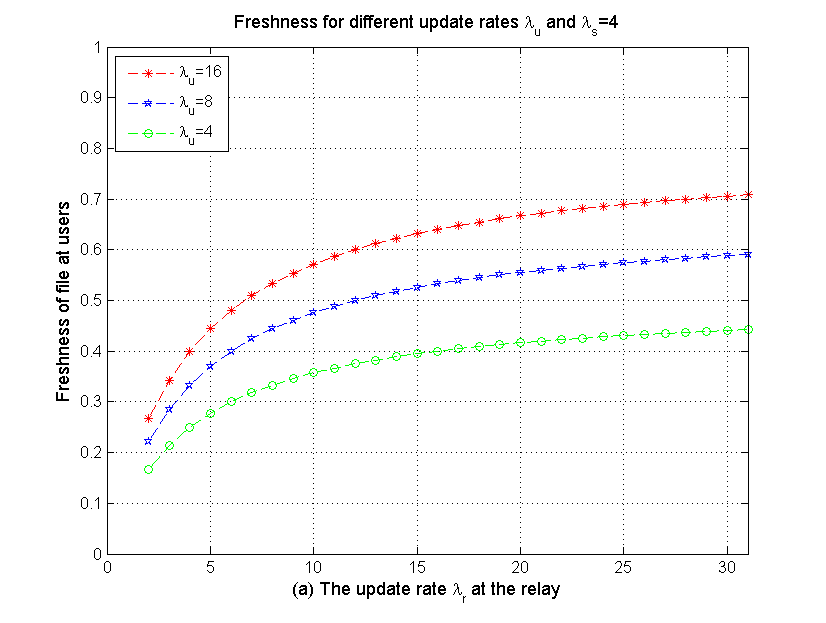}
\includegraphics[width=3in]{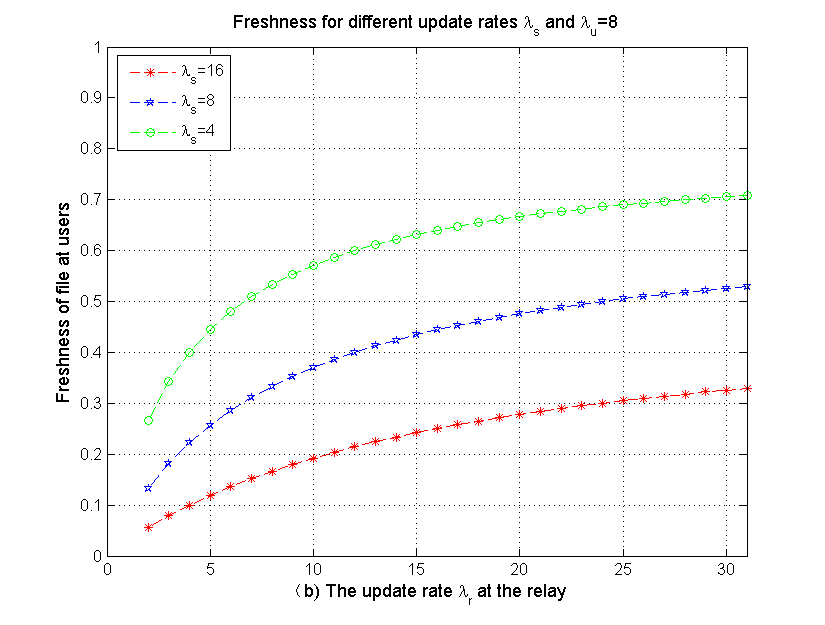}
\vskip 0.2cm
\caption{The relationship between freshness and the update rate at the relay, user, server}\label{figure2}
\end{figure}
As shown in Fig.\ref{figure2}, $F_{r_{k},u_{i},f_{j}}$ is an increasing concave function of the update rate $\lambda_{r_{k},f_{j}}$. For the same freshness $F_{r_{k},u_{i},f_{j}}$, with the increase of the update rate $\lambda_{u_{i},f_{j}}$ at the user, the update rate $\lambda_{r_{k},f_{j}}$ becomes smaller significantly (see $Fig.2(a)$). However, with the increase of the update rate $\lambda_{r_{0},f_{j}}$ at the server, the update rate $\lambda_{r_{k},f_{j}}$  becomes larger significantly (see $Fig.2(b)$). This implies the optimal update rate allocation policy for the relay can be studied according the number of the files requested by different users. So the update rates $\lambda_{r_{k},f_{j}}$ are uncertain integer variables in AFOCUP problem. We rewrite the optimal model based on \eqref{eq-maxim-7} specifically as follows
\begin{eqnarray}
&\max& F=\frac{1}{M}\sum_{i=1}^{M}F_{u_{i}}  \notag\\
& &~~=\frac{1}{M}\sum_{i=1}^{M}\sum_{f_{j}\in\mathcal{F}_{u_{i}}}p_{u_{i},f_{j}}\sum_{k=1}^{K}p_{r_{k},u_{i}}\cdot \frac{\lambda_{u_{i},f_{j}}}{\lambda_{u_{i},f_{j}}+\lambda_{r_{0},f_{j}}}\cdot \frac{\lambda_{r_{k},f_{j}}}{\lambda_{r_{k},f_{j}}+\lambda_{r_{0},f_{j}}}\cdot x_{r_{k},u_{i},f_{j}}.\label{eq-maxim-8}\\
&s.t.& conditions~(2),~(3),~(4),~(5),  \notag \\
& & x_{r_{k},u_{i},f_{j}}\in X , ~\lambda_{r_{k},f_{j}} \in Z.
\end{eqnarray}

Obviously, the above model is nonlinear integer programming problem. In the following, we provide an approximation of the AFOCUP problem to seek out the optimal cache scheme.

\subsection{Upper Bound Approximation To AFOCUP}
Due to the AFOCUP's intractability, it is difficult to obtain the optimal integer solution. For problem-solving, we linearize both the objective function and the first inequality constraints. These together give us a 0-1 linear integer programming model, as an approximation to the original model. The intractability of AFOCUP lies in the product of variables $x_{r_{k},u_{i},f_{j}}$ and $\lambda_{r_{k},f_{j}}$ contained in the objective function \eqref{eq-maxim-8} and the first constraint condition \eqref{eq-maxim-2}, which is the cause for nonlinearity. Overcoming this difficulty potentially motivates us to find an effective approach to make an approximation of the original model.

In order to utilize the total update rate capacity sufficiently at the relay, firstly, we replace the first inequality constraint \eqref{eq-maxim-2} with the following equality constraint
\[
\sum_{i=1}^{M}\sum_{j=1}^{n_{i}}\lambda_{r_{k},f_{j}}\cdot x_{r_{k},u_{i},f_{j}}=G_{k},~k=1,2,\ldots,K.
\]

For an arbitrary cache scheme $X$ satisfying with the rest constraints \eqref{eq-maxim-3}, \eqref{eq-maxim-4}, \eqref{eq-maxim-5}, we can obtain the specific number $T_{k}$ of files cached in relay $r_{k}$, i.e., $T_{k}\leq C_{k}$. However, the update rates $\lambda_{r_{k},f_{j}}$ are still uncertain even though $\lambda_{u_{i},f_{j}}$ and $\lambda_{r_{0},f_{j}}$ are known for the relay $r_{k}$. As illustrated in Fig.2, there should be an optimal update rate allocation for each relay $r_{k}$ based on the certain cache scheme $X$. Thus, an optimal model is proposed by distributing the update rate $\lambda_{r_{k},f_{j}}$ completely in accordance with the total update rate $G_{k}$ and file number $T_{k}$ stored in the relay $r_{k}$. Specifically, in order to obtain the maximum freshness of these files at the users, we establish an optimal update rate allocation problem with regard to all relays $r_{k}$, $(k=1,2,\ldots,K)$ as follows
\begin{eqnarray}
&\max\limits_{\lambda_{r_{k},u_{i},f_{j}}}&~\sum_{j=1}^{T_{k}}
\frac{\lambda_{u_{i},f_{j}}}{\lambda_{u_{i},f_{j}}+\lambda_{r_{0},f_{j}}}\cdot
\frac{\lambda_{r_{k},f_{j}}}{
\lambda_{r_{k},f_{j}}+\lambda_{r_{0},f_{j}}}  \label{eq-maxim-10}\\
&s.t.& ~\sum_{j=1}^{T_{k}}\lambda_{r_{k},f_{j}}\leq G_{k}, \notag\\
&    & ~\lambda_{r_{k},f_{j}} \geq 0, ~~~~j=1,2,\ldots,T_{k}. \notag
\end{eqnarray}

 By solving the model \eqref{eq-maxim-10}, we can obtain an analytical expression for the update rate $\lambda_{r_{k},f_{j}}$, which is represented by the update rates $\lambda_{r_{0},f_{j}}$ and $\lambda_{r_{u_{i}},f_{j}}$. With the help of the optimal expression, an upper bound approximation of AFOCUP is provided as the following theorem.
\begin{theorem}
\label{th-1}
For file $f_{j}$ at user $u_{i}$,  when the update rates $\lambda_{r_{k},f_{j}}$ at the relay $r_{k}$ is subject to the condition
\begin{eqnarray}
\lambda_{r_{k},f_{j}}^{*}=max\left\{
\frac{\beta}{\alpha}
\sqrt{\frac{\lambda_{u_{i},f_{j}}\cdot\lambda_{r_{0},f_{j}}}{\lambda_{u_{i},f_{j}}+\lambda_{r_{0},f_{j}}}
}-\lambda_{r_{0},f_{j}} , 0\right\} \label{eq-maxim-11}
\end{eqnarray}
\noindent where
$\alpha=\sum_{j=1}^{T_{k}}\sqrt{
\frac{\lambda_{u_{i},f_{j}}\cdot\lambda_{r_{0},f_{j}}}
{\lambda_{u_{i},f_{j}}+\lambda_{r_{0},f_{j}}}
}$,~
$\beta=G_{k}+\sum_{j=1}^{T_{k}}\lambda_{r_{0},f_{j}}$.
then, the optimal solution of the 0-1 integer linear programming model as follows
\begin{eqnarray}
&\max& FR^{*}=\frac{1}{M}\sum_{i=1}^{M}FR_{u_{i}}^{*}
=\frac{1}{M}\sum_{i=1}^{M}\sum_{f_{j}\in\mathcal{F}_{u_{i}}}p_{u_{i},f_{j}}\sum_{k=1}^{K}
p_{r_{k},u_{i}}\cdot F_{r_{k},u_{i},f_{j}}^{*}\cdot x_{r_{k},u_{i},f_{j}},  \label{eq-maxim-12}\\
&s.t.& \sum_{i=1}^{M}\sum_{f_{j}\in\mathcal{F}_{u_{i}}}x_{r_{k},u_{i},f_{j}}\leq C_{k},~~~k=1,2,\ldots,K, \notag\\
&    & \sum_{k=1}^{K}x_{r_{k},u_{i},f_{j}}=1,~~~~~~~~~~~~~u_{i}\in U, f_{j}\in\mathcal{F}_{u_{i}}, \notag\\
&    & \sum_{f_{j}\in\mathcal{F}_{u_{i}}}\sum_{k=1}^{K}x_{r_{k},u_{i},f_{j}}=n_{i},~~~~u_{i}\in U, \notag \\
&    &  x_{r_{k},u_{i},f_{j}}\in X.  \notag
\end{eqnarray}
\noindent where $
F_{r_{k},u_{i},j}^{*}=\frac{
\lambda_{u_{i},f_{j}}}{
\lambda_{u_{i},f_{j}}+\lambda_{r_{0},f_{j}}
}\cdot\frac{
\lambda_{r_{k},f_{j}}^{*}
}{
\lambda_{r_{k},f_{j}}^{*}+\lambda_{r_{0},f_{j}}
}$,
is upper bound approximation of AFOCUP.
\end{theorem}
The proof of Theorem \ref{th-1} is attached in Appendix A. In order to obtain the optimal cache scheme of the AFOCUP problem, we will provide two algorithms based on the optimal model  \eqref{eq-maxim-10} and  \eqref{eq-maxim-11}. Algorithm 1 is used to get the optimal update rate distribution of all files in each relay for an arbitrary specific cache scheme. Based on the facts that the higher of average freshness of all files at the users, an improved the cache scheme realized by Algorithm 2 is obtained to search for the optimal cache scheme with the aid of Algorithm 1.

For convenience's sake, The Algorithm 1 is named as Relay-Optimal Request Update Rate Policy Under Total Update Rate Constraint Algorithm (RUPA) as follows

\begin{center}
\begin{algorithm}[H]
\caption{\textbf{(RUPA)}Relay-Optimal Request Update Rate Policy Under Total Update Rate Constraint}
\LinesNumbered
\KwIn{
$G_{k}$-the total update rates of $r_{k}$;
$T_{k}$-the number of files cached in $r_{k}$ in specific cache scheme $X$;
$F_{r_{k}}$-the files set cached in relay $r_{k}$;
$\lambda_{r_{0},F_{r_{k}}}$-the update rate set of $F_{r_{k}}$ at the source;
$\lambda_{u,F_{r_{k}}}$-the update rate set of $F_{r_{k}}$ at the user ;
}
\KwOut{
$\lambda_{r_{k},F_{r_{k}}}$-the vector specifying optimal the request update rate for each file $f_{j}\in F_{r_{k}}$ under the condition $G_{k}$.
}
Let $\mu_{j}\leftarrow\frac{\lambda_{u,f_{j}}}{\lambda_{u,f_{j}}+\lambda_{r_{0},f_{j}}}$;
    $\alpha\leftarrow\sum_{j=1}^{T_{k}}\sqrt{\mu_{j}\cdot\lambda_{r_{0},f_{j}}}$;
    $\beta\leftarrow\sum_{j=1}^{T_{k}}\lambda_{r_{0},f_{j}}+G_{k}$\;
Sort $\mu_{j}$, $\lambda_{r_{0},f_{j}}$ in non-descending order of $\mu_{j}/\lambda_{r_{0},f_{j}}$ for file $f_{j}$\;
\For {~j=1 to $T_{k}$~}{
      \eIf{~$\mu_{j}/\lambda_{r_{0},f_{j}}\leq \frac{\alpha^{2}}{\beta^{2}}$~}{
     $\lambda_{r_{k},f_{j}}\leftarrow 0$\;
     $\alpha\leftarrow\alpha-\sqrt{\mu_{j}\cdot\lambda_{r_{0},f_{j}}}$\;
     $\beta\leftarrow\beta-\lambda_{r_{0},f_{j}}$\;
     }
     {
     $\lambda_{r_{k},f_{j}}\leftarrow \beta\sqrt{\mu_{j}\cdot\lambda_{r_{0},f_{j}}}/\alpha-\lambda_{r_{0},f_{j}}$\;
     }
}
Return $\lambda_{r_{k},F_{r_{k}}}$
\end{algorithm}
\end{center}
\begin{center}
\begin{algorithm}[H]
\caption{The Greedy Algorithm to search for the optimal cache scheme with the help of algorithm 1\textbf{(RUPA)}}
\LinesNumbered
\KwIn{
$PS$-the permutation set of $N$ files;~~$CS$-the feasible partition set of $N$ files\;
~~~~~~~~~~$PR$-the probability set of all relays requested by all users\;
~~~~~~~~~~$PU$-the probability set of Files requested by users;
}
\KwOut{
$CST$-the optimal cache scheme information table\;
~~~~~~~~~~~~$Fresh$-the optimal average fresh of the cache updating system.
}
Initialize the data: $OptimalCST=0; OptimalFresh=0; FRESH=0;Relay=K; File=N; User=M$\;
\For{i=1 to $LPS$}{
\For{j=1 to $LCS$}{
$n\leftarrow CS$\;
$RC\leftarrow G$\;
$UU\leftarrow \{\{\lambda_{r_{c_{i}},f,u}\}:c_{i}\in CS, i=1,\ldots,K\}$\;
$SU\leftarrow \{\{\lambda_{s_{c_{i}},f,u}\}:c_{i}\in CS, i=1,\ldots,K\}$\;
\For{~r=1 to $K$~}{
$RU\leftarrow \textbf{RUPA}(n,RC,UU,SU)$
}\
$cst\leftarrow (FI,UI,UU,RI,RU,SU)_{N\times6}$\;
$Fresh\leftarrow \sum_{i=1}^{N}\frac{cst_{i,3}}{(cst_{i,3}+cst_{i,6})}\cdot\frac{cst_{i,5}}{(cst_{i,5}+cst_{i,6})}\cdot        
PU_{cst_{i,2},cst_{i,1}}\cdot PR_{cst_{i,2},cst_{i,4}}$\;
$tmp\leftarrow Fresh$\;
\eIf{$tmp>FRESH(i)$}{
$FRESH(i)\leftarrow tmp$\;
$CST(i)\leftarrow cst$
}{
$FRESH(i)\leftarrow FRESH(i)$\;
$CST(i)\leftarrow CST(i)$\;
}
}\
$tmp\leftarrow FRESH(i)$\;
\eIf{$tmp>OptimalFresh$}{
$OptimalFresh\leftarrow tmp$\;
$OptimalCST\leftarrow CST(i)$\;
}{
$OptimalFresh\leftarrow OptimalFresh$\;
$OptimalCST\leftarrow OptimalCST$\;
}
}
Return~~$OptimalFresh$;~~$OptimalCST$
\end{algorithm}
\end{center}
For a given cache scheme $X$ satisfied with constraint \eqref{eq-maxim-3}, \eqref{eq-maxim-4}, \eqref{eq-maxim-5}, \eqref{eq-maxim-6}, Algorithm 1 (RUPA) is used to obtain the optimal update rate $\lambda_{r_{k},f_{j}}$ according to the formula \eqref{eq-maxim-11}. Then we utilize the objective function \eqref{eq-maxim-12} to get the average freshness of cache scheme $X$. If the average freshness is larger than that before, we will replace the previous one with the current cache scheme. Therefore, the Greedy Algorithm can be used to search for the optimal cache scheme in the Algorithm 2.

For convenience, we make a notation on Algorithm 2. $PS$ is a set of all permutations of file index, i.e.
\begin{eqnarray*}
PS=\{(p_{1},p_{2},\ldots,p_{N}): p_{i}\in\{1,\ldots,N\},\ ~\forall p_{i}\neq p_{j}\}
\end{eqnarray*}
$LPS$ is the length of $PS$, i.e. $|PS|=N!$.

$CS$ is a set of the feasible partition for $N$ files based on the cache capacity $C_{k}$ of Relay $r_{k}$, i.e.
\begin{eqnarray*}
CS=\left\{(c_{1},c_{2},\ldots,c_{K}):\sum_{i=1}^{K}c_{i}=N,c_{i}\in Z^{+}\right\}
\end{eqnarray*}
$LCS$ is the length of $CS$.

$CST$ is the cache scheme information table, which is a matrix with $N$ rows and 6 columns and covers the information such as File indexes, User indexes, Update rates at users, Relay indexes, Update rates at relays, Update rates at server.(see TABLE II)

For an permutation contained $N$ number, the $N$ files at the server are ranked as the given permutation and divided into $K$ subfile sets based on the element of $CS$ set. Each subfile set is identified as the cache scheme at the relays. In addition, an given permutation means an storage status at the users. Therefore, we can search for the optimal partition $ (c_{1},c_{2},\ldots,c_{K})$ by using the greedy algorithm to calculate the average freshness with the aid of Algorithm 1.

\section{NUMERICAL RESULTS}
In this section, simulation results will be provided to evaluate the performance of the proposed approximation of the AFOCUP problem. Moreover, we will validate the AFOCUP and the effectiveness of the greedy caching strategy. In the following, we assume that the file request probability follows a Zipf distribution with parameter $\alpha$ in \cite{BPLPS}, i.e.
\begin{eqnarray*}
p_{f}=\frac{f^{-\alpha}}{\sum_{i\in\mathcal{F}}i^{-\alpha}}.
\end{eqnarray*}

For the first example, the main parameters are designed as File number $N=10$, Relay number $K=3$, User number $M=4$. The other  data information is given as TABLE I.

\begin{center}
\begin{table}[h]
\centering
\caption{Simulation Parameters Of Caching Update System}\label{tab:aStrangeTable}
\begin{tabular}{cccc}
\toprule  
Parameters~~~~~~~~~~~~~~~~~~~~~~~~~~~~~~~~~~~& Value       &Parameters~~~~~~~~~~~~~~~~~~~~~~~~~~~~~~~~~~~& Value\\
\midrule  
File number~~~~~~~~~~~~~~~~~~~~~~~~~~~~~~~~~~& 10          &User number~~~~~~~~~~~~~~~~~~~~~~~~~~~~~~~~~~& 4 \\
Relay number~~~~~~~~~~~~~~~~~~~~~~~~~~~~~~~~~& 3           &Update rate of all files at server~~~~~~~~~~~& (4,3,3,6,4,3,6,4,5,6)\\
Cache capability of relays~~~~~~~~~~~~~~~~~~~& (6,5,4)     &Total Update rates of relays~~~~~~~~~~~~~~~~~&(12,10,8)  \\
File index at User 1~~~~~~~~~~~~~~~~~~~~~~~~~& (1,2,3)     &File index at User 2~~~~~~~~~~~~~~~~~~~~~~~~~& (4,5,6)\\
File index at User 3~~~~~~~~~~~~~~~~~~~~~~~~~& (7,8)       &File index at User 4~~~~~~~~~~~~~~~~~~~~~~~~~& (9,10)\\
Update rate of files at User 1~~~~~~~~~~~~~~~& (8,10,12)   &Update rate of files at User 2~~~~~~~~~~~~~~~& (6,12,8)\\
Update rate of files at User 3~~~~~~~~~~~~~~~& (10,10)     &Update rate of files at User 4~~~~~~~~~~~~~~~& (12,6)\\
Probability of files requested by User 1~~~& (0.3,0.3,0.4) &Probability of files requested by User 2~~~& (0.2,0.3,0.5)\\
Probability of files requested by User 3~~~& (0.4,0.6)     &Probability of files requested by User 4~~~& (0.5,0.5)\\
Probability of relays requested by User 1 & (0.5,0.3,0.2)  &Probability of relays requested by User 2 & (0.3,0.5,0.2)\\
Probability of relays requested by User 3 & (0.2,0.5,0.3)  &Probability of relays requested by User 4 & (0.4,0.3,0.3)\\
\bottomrule 
\end{tabular}
\end{table}
\end{center}

According to Algorithm 1 and Algorithm 2, we can obtain the optimal cache scheme information table $CST$ as TABLE II. At the same time, we also provide the increasing tendency of freshness during the iteration. (see Fig.3)

In practice, the average freshness is related to the popularity of requested files in the cache updating system. The method of freshness calculation in [23] suggested that the popularity of all files in the server is consistent. However, the method proposed in this paper considers the discrepancy of file popularity in cache updating system. The following example is used to demonstrate the performance in the case of different file popularity.
\begin{eqnarray*}
\\
\end{eqnarray*}
\begin{center}
\begin{table}[h]
\centering
\caption{The Optimal Cache Scheme Data Table}\label{tab:aStrangeTable}
\begin{tabular}{ccccccccccc}
\toprule  
File Index & User Index & Update rate at User & Relay Index &  Update rate at Relay & Update rate at Server \\
\midrule  
1 & 1 & 8  & 1 & 2.4832 & 4 \\
2 & 1 & 10 & 1 & 3.0311 & 3 \\
3 & 1 & 12 & 1 & 3.1505 & 3 \\
4 & 2 & 6  & 1 & 0.8765 & 6 \\
5 & 2 & 12 & 2 & 3.4239 & 4 \\
6 & 2 & 8  & 2 & 3.3311 & 3 \\
7 & 3 & 10 & 3 & 4.5573 & 6 \\
8 & 3 & 10 & 2 & 3.2450 & 4 \\
9 & 4 & 12 & 1 & 2.4586 & 5 \\
10& 4 & 6  & 3 & 3.4427 & 6 \\
\midrule  
Users& 4 & Relays& 3 & Files& 10\\
\midrule
Optimal Fresh & & &0.5319&&\\
\bottomrule
\end{tabular}
\end{table}
\end{center}

\begin{figure}[h]
\centering
\includegraphics[width=3in]{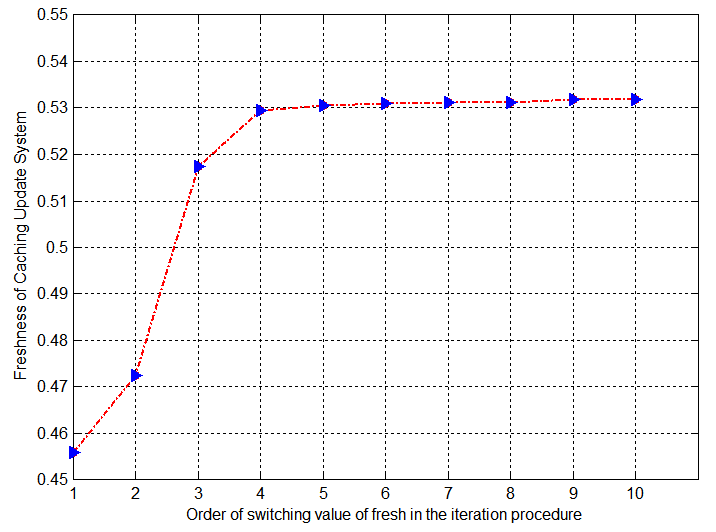}
\vskip 0.2cm
\caption{The increasing tendency of freshness}\label{figure3}
\end{figure}
\
\begin{figure}[h]
\centering
\includegraphics[width=3in]{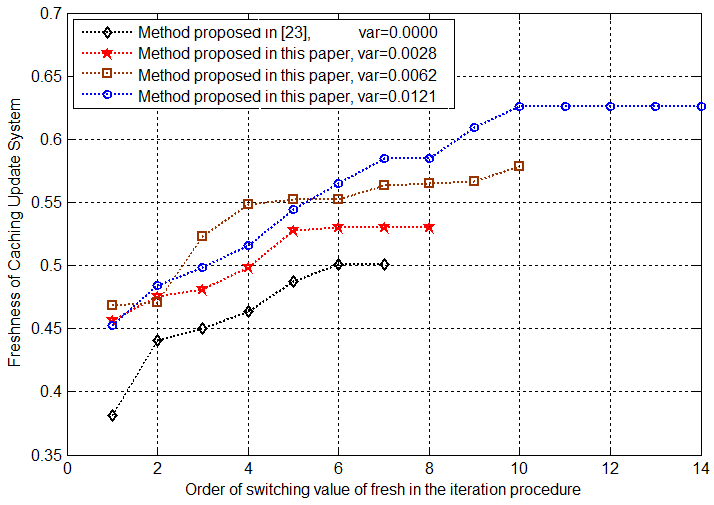}
\vskip 0.2cm
\caption{A comparison of the performance of the optimal solution with respect to inconsistent file popularity with the same mean 0.1 and different variance }\label{figure4}
\end{figure}

For the second example, we provide four different sets of data on file popularity with the same mean and different variance. According to Algorithm 1 and Algorithm 2, we can obtain the values of the optimal average freshness are 0.5012, 0.5305, 0.5783 and 0.6263. The increasing tendency of freshness during the iteration are shown in the Fig.4. Fig.4 shows that the optimal solution of the average freshness in the cache updating system is monotonically increasing on the variance of values of file popularity, which implied that the file with much more popularity can be updated prior to all other files in the relays.

In addition, when the update rates of files at users or server change, the freshness of the cache updating system will change accordingly.  In order to better explain the specific relationship between the freshness and update rates at users or server,
we provide other two examples as follows

\begin{center}
\begin{table}[h]
\centering
\caption{Three different data sets of update rates at users}\label{tab:aStrangeTable}
\begin{tabular}{ccccccccccc}
\toprule  
File Index & User Index & Data sets 1 & Data sets 2 &  Data sets 3 \\
\midrule  
1 & 1 & 4 & 8  & 12 \\
2 & 1 & 5 & 10 & 15 \\
3 & 1 & 6 & 12 & 18 \\
4 & 2 & 3 & 6  & 9  \\
5 & 2 & 6 & 12 & 18 \\
6 & 2 & 4 & 8  & 12 \\
7 & 3 & 5 & 10 & 15 \\
8 & 3 & 5 & 10 & 15 \\
9 & 4 & 6 & 12 & 18 \\
10& 4 & 3 & 6  & 9  \\
\bottomrule 
\end{tabular}
\end{table}
\end{center}

\begin{figure}[h]
\centering
\includegraphics[width=3in]{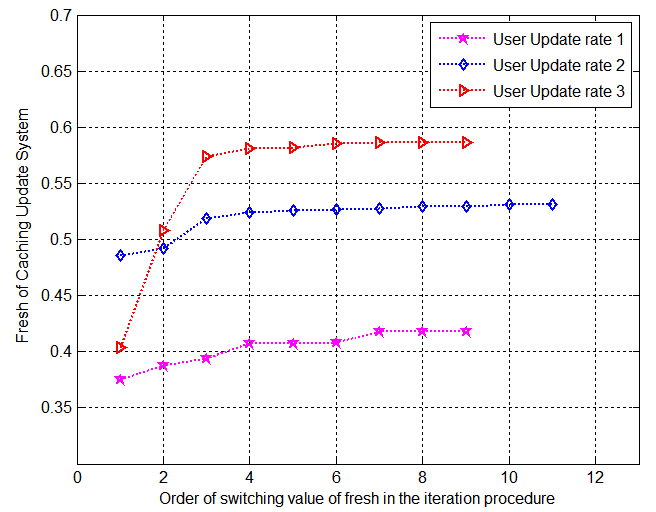}
\vskip 0.2cm
\caption{The increasing tendency of freshness corresponding to TABLE III}\label{figure5}
\end{figure}

For the third example, we provide three data sets with different file update rates at users and the same file update rates at the server ($see$ TABLE III). By using Algorithm 1 and Algorithm 2, we can get the corresponding optimal freshness as follows 0.4184, 0.5139 and 0.5861. The increasing tendency of freshness corresponding to TABLE III during the iteration are shown as Fig.5. From the result of calculations, we find that the freshness of the cache updating system is monotonically increasing on file update rates at users.

\begin{center}
\begin{table}[h]
\centering
\caption{Three different data sets of update rates at the server}\label{tab:aStrangeTable}
\begin{tabular}{ccccccccccc}
\toprule  
File Index & Data sets 1 & Data sets 2 &  Data sets 3 \\
\midrule  
1 & 4 & 6 & 8  \\
2 & 3 & 4 & 6  \\
3 & 3 & 4 & 6  \\
4 & 6 & 8 & 10 \\
5 & 4 & 6 & 8  \\
6 & 3 & 4 & 6  \\
7 & 6 & 8 & 10 \\
8 & 4 & 6 & 8  \\
9 & 5 & 7 & 9  \\
10& 6 & 8 & 10 \\
\bottomrule 
\end{tabular}
\end{table}
\end{center}
\
\begin{figure}[h]
\centering
\includegraphics[width=3in]{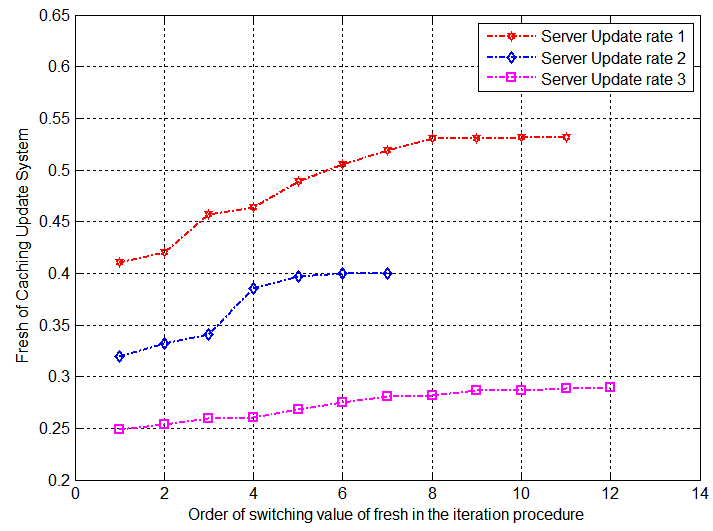}
\vskip 0.2cm
\caption{The increasing tendency of freshness corresponding to TABLE IV}\label{figure6}
\end{figure}

For the four example, we provide three data sets with different file update rates at server and the same file update rates at users($see$ TABLE IV). By using Algorithm 1 and Algorithm 2, we can get the corresponding optimal freshness as follows 0.5139, 0.3998 and 0.2894. The increasing tendency of freshness corresponding to TABLE IV sets during the iteration are shown in the Fig.6. From the result of calculations, we find that the optimal freshness of the cache updating system is monotonically decreasing on file update rates at the server.

\section{Conclusion}   
\label{conclusion}
In this paper, we considered a cache updating system with a server, $K$ relays and $M$ users. There are $N$ files in the server. Each file is updated by the server with a fixed rate. Each relay have a limited storage capacity as well as a total update rates. Each user can update a file by any relay, but the precondition is the file must be cached in the requested relay. We defined a cache updating scheme and then proposed an average freshness-optimal caching updating problem to obtain the optimal cache scheme. Because of the proposed optimal model being nonlinear integer programming problem, it is difficult to obtain the optimal solution. As a result, we provided an linear approximate model to replace the original model. With the help of the greedy algorithm, we can solve the 0-1 linear integer programming model and search for the optimal cache scheme. In addition, through specific numerical data analysis, we found that the optimal average freshness of the cache updating system is monotonically decreasing on file update rates at the server and monotonically increasing on file update rates at users.

\section*{Appendix A: The proof of Theorem \ref{th-1}}
\begin{proof}
For the above optimal update rate allocation problem \eqref{eq-maxim-10}, we introduce the Lagrangian function for variable constraints $\lambda_{r_{k},f_{j}} \geq 0$ as
\begin{eqnarray*}
\textsl{L}=-\sum_{j=1}^{T_{k}}
\frac{
\lambda_{u_{i},f_{j}}
}{
\lambda_{u_{i},f_{j}}+\lambda_{r_{0},f_{j}}
}\cdot
\frac{
\lambda_{r_{k},f_{j}}
}{
\lambda_{r_{k},f_{j}}+\lambda_{r_{0},f_{j}}
}
+\delta\cdot\left(\sum_{j=1}^{T_{k}}\lambda_{r_{k},f_{j}}-G_{k}\right)
-\left(\sum_{j=1}^{T_{k}}\theta_{j}\cdot\lambda_{r_{k},f_{j}}\right)
\end{eqnarray*}
where $\delta\geq0, \theta_{j}\geq0$. Then we write the $KKT$ conditions as
\begin{eqnarray}
\frac{
\partial L
}{
\partial\lambda_{r_{k},f_{j}}
}
=-\frac{
\lambda_{u_{i},f_{j}}
}{
\lambda_{u_{i},f_{j}}+\lambda_{r_{0},f_{j}}
}\cdot
\frac{
\lambda_{r_{0},f_{j}}
}{
(\lambda_{r_{k},f_{j}}+\lambda_{r_{0},f_{j}})^{2}
}
+\delta-\theta_{j}=0, \label{eq-maxim-13}
\end{eqnarray}
for all $j$. The complementary slackness conditions are
\begin{eqnarray}
\delta\cdot\left(\sum_{j=1}^{T_{k}}\lambda_{r_{k},f_{j}}-G_{k}\right)&=&0,\label{eq-maxim-14}\\
\theta_{j}\cdot\lambda_{r_{k},f_{j}}&=&0.\label{eq-maxim-15}
\end{eqnarray}

If $\lambda_{r_{k},f_{j}}>0$, we have $\theta_{j}=0$ from \eqref{eq-maxim-15}, Thus, we rewrite \eqref{eq-maxim-13} as
\begin{eqnarray*}
(\lambda_{r_{k},f_{j}}+\lambda_{r_{0},f_{j}})^{2}=
\frac{1}{\delta}
\frac{\lambda_{u_{i},f_{j}}\cdot\lambda_{r_{0},f_{j}}}{\lambda_{u_{i},f_{j}}+\lambda_{r_{0},f_{j}}}
\end{eqnarray*}
and then get
\begin{eqnarray}
\lambda_{r_{k},f_{j}}=max \{
\frac{1}{\sqrt{\delta}}
\sqrt{\frac{\lambda_{u_{i},f_{j}}\cdot\lambda_{r_{0},f_{j}}}{\lambda_{u_{i},f_{j}}+\lambda_{r_{0},f_{j}}}}
-\lambda_{r_{0},f_{j}}, 0\}.\label{eq-maxim-16}
\end{eqnarray}
with regard to \eqref{eq-maxim-14}, we can get
\begin{eqnarray*}
\delta=\frac{\left(\sum_{j=1}^{T_{k}}\sqrt{\frac{\lambda_{u_{i},f_{j}}\cdot\lambda_{r_{0},f_{j}}}
{\lambda_{u_{i},f_{j}}+\lambda_{r_{0},f_{j}}}
}\right)^{2}}
{(G_{k}+\sum_{j=1}^{T_{k}}\lambda_{r_{0},f_{j}})^{2}}
\end{eqnarray*}
Let~ $\alpha=\sum_{j=1}^{T_{k}}\sqrt{\frac{\lambda_{u_{i},f_{j}}\cdot\lambda_{r_{0},f_{j}}}
{\lambda_{u_{i},f_{j}}+\lambda_{r_{0},f_{j}}}}$,
~$\beta=G_{k}+\sum_{j=1}^{T_{k}}\lambda_{r_{0},f_{j}}$,
Then, $\frac{1}{\sqrt{\delta}}=\frac{\beta}{\alpha}$,
Thus, we can rewrite \eqref{eq-maxim-16} as
\begin{eqnarray*}
\lambda_{r_{k},f_{j}}^{*}=\max \left\{
\frac{\beta}{\alpha}
\sqrt{\frac{\lambda_{u_{i},f_{j}}\cdot\lambda_{r_{0},f_{j}}}{\lambda_{u_{i},f_{j}}+\lambda_{r_{0},f_{j}}}}
-\lambda_{r_{0},f_{j}} , 0\right\}.
\end{eqnarray*}

For given update rates $\lambda_{r_{0},f_{j}}$ at the server, the above relationship between $\lambda_{r_{k},f_{j}}$ and $\lambda_{u_{i},f_{j}}$ can ensure the maximum freshness of $T_{k}$ files at the relay $r_{k}$. As a result, we rewrite the objective function \eqref{eq-maxim-8} as
\begin{eqnarray*}
F&=&\frac{1}{M}\sum_{i=1}^{M}\sum_{f_{j}\in\mathcal{F}_{u_{i}}}p_{u_{i},f_{j}}
\sum_{k=1}^{K}p_{r_{k},u_{i}}\cdot F_{r_{k},u_{i},f_{j}}\cdot x_{r_{k},u_{i},f_{j}}\\
&=&\frac{1}{M}\sum_{k=1}^{K}p_{r_{k},u_{i}}\sum_{i=1}^{M}\sum_{f_{j}\in\mathcal{F}_{u_{i}}}p_{u_{i},f_{j}}
F_{r_{k},u_{i},f_{j}}\cdot x_{r_{k},u_{i},f_{j}}
\end{eqnarray*}
For relay $r_{k}$, we have $\sum_{f_{j}\in\mathcal{F}_{u_{i}}}\sum_{i=1}^{M}x_{r_{k},u_{i},f_{j}}=T_{k}\leq C_{k}$.
When the update rate $\lambda_{r_{k},f_{j}}$ is replaced with $\lambda_{r_{k},f_{j}}^{*}$ in the formula $F_{r_{k},u_{i},f_{j}}$, we can obtain the inequality
\begin{eqnarray*}
\sum_{i=1}^{M}\sum_{f_{j}\in\mathcal{F}_{u_{i}}}p_{u_{i},f_{j}}F_{r_{k},u_{i},f_{j}}x_{r_{k},u_{i},f_{j}}
\leq\sum_{i=1}^{M}\sum_{f_{j}\in\mathcal{F}_{u_{i}}}p_{u_{i},f_{j}}F_{r_{k},u_{i},f_{j}}^{*}x_{r_{k},u_{i},f_{j}},
\end{eqnarray*}
Thus, on the one hand, the objective function $F$ from \eqref{eq-maxim-8} is subject to
\begin{eqnarray*}
F &\leq & FR^{*}\\
&=& \frac{1}{M}\sum_{i=1}^{M}FR_{u_{i}}^{*}
=\frac{1}{M}\sum_{i=1}^{M}\sum_{f_{j}\in\mathcal{F}_{u_{i}}}p_{u_{i},f_{j}}
\sum_{k=1}^{K}p_{r_{k},u_{i}}\cdot F_{r_{k},u_{i},f_{j}}^{*}\cdot x_{r_{k},u_{i},f_{j}},
\end{eqnarray*}
On the other hand, for the constraint \eqref{eq-maxim-2}, the above equation \eqref{eq-maxim-14} shows $\sum_{j=1}^{T_{k}}\lambda_{r_{k},f_{j}}^{*}=G_{k}$, it is apparent for relay $r_{k}$ that
\begin{eqnarray*}
\sum_{f_{j}\in\mathcal{F}_{u_{i}}}\sum_{i=1}^{M}\lambda_{r_{k},f_{j}}\cdot x_{r_{k},u_{i},f_{j}}
\leq \sum_{j=1}^{T_{k}}\lambda_{r_{k},f_{j}}^{*}=G_{k},~~k=1,2,\ldots,K.
\end{eqnarray*}
Thus, the optimal solution of 0-1 integer linear programming model \eqref{eq-maxim-12} is upper bound approximation of AFOCUP.
\end{proof}


\begin{thebibliography}{1}

\bibitem{SGDMC}
K. Shanmugam, N.Golrezaei, A.G. Dimakis, A.F. Molisch, and G.Caire. FemtoCaching: Wireless content delivery through distributed caching helpers. IEEE Trans. Inf. Theory, vol. 59, no. 12, pp. 8402-8413, Dec.2013.

\bibitem{WCTKL}
X. Wang, M. Chen, T. Taleb, A. Ksentini, and V. C. M. Leung. Cache in the air: Exploiting content caching and delivery techniques for 5G systems. IEEE Commun. Mag., vol.52, no.2, pp.131-139, Feb. 2014.

\bibitem{CJW}
Y. Cui, D. Jiang, and Y. Wu. Analysis and optimization of caching and multicasting in large-scale cache-enabled wireless networks. IEEE Trans. Wireless Commun., vol.16, no.1, pp.250-264, Jan. 2017.

\bibitem{LWZZY}
J. Liao, K.-K. Wong, Y. Zhang, Z. Zheng, and K. Yang. Coding, multicast, and cooperation for cache-enabled heterogeneous small cell networks. IEEE Trans. Wireless Commun., vol.16, no.10, pp.6838-6853, Oct. 2017.

\bibitem{DAFY1}
T. Deng, G. Ahani, P. Fan, and D. Yuan. Cost-optimal caching for D2D networks with presence of user mobility. IEEE GLOBECOM, pp.1-6, Dec. 2017.

\bibitem{DAFY2}
T. Deng, G. Ahani, P. Fan, and D. Yuan. Cost-optimal caching for D2D networks with user mobility: modeling, analysis, and
computational approaches. IEEE Trans. Wireless Commun., vol.17, no.5, pp.3082-3094, May 2018.

\bibitem{ICM}
S. Ioannidis, A. Chaintreau, and L. Massoulie. Optimal and scalable distribution of content updates over a mobile social network. In IEEE Infocom, April 2009.

\bibitem{CG}
J. Cho and H. Garcia-Molina. Effective page refresh policies for web crawlers. ACM Transactions on Database Systems,
vol. 28, no. 4, pp. 390-426, Dec. 2003.

\bibitem{AHLPS}
Y. Azar, E. Horvitz, E. Lubetzky, Y. Peres, and D. Shahaf. Tractable near-optimal policies for crawling. PNAS, vol. 115, no. 32, pp.8099-8103, August 2018.

\bibitem{KPLH}
A. Kolobov, Y. Peres, E. Lubetzky, and E. Horvitz. Optimal freshness crawl under politeness constraints. In ACM SIGIR Conference, pp.495-504, July 2019.

\bibitem{BSS}
A. M. Bedewy, Y. Sun, and N. B. Shroff. Optimizing data freshness, throughput, and delay in multi-server information-update systems. In IEEE ISIT, July 2016.

\bibitem{HYE}
Q. He, D. Yuan, and A. Ephremides. Optimizing freshness of information: On minimum age link scheduling in wireless
systems. In IEEE WiOpt, May 2016.

\bibitem{SUY}
Y. Sun, E. Uysal-Biyikoglu, R. D. Yates, C. E. Koksal, and N. B. Shroff. Update or wait: How to keep your data fresh.
IEEE Trans. Inf. Theory, vol. 63, no. 11, pp.7492-7508, November 2017.

\bibitem{SU}
A. Soysal and S. Ulukus. Age of information in G/G/1/1 systems: Age expressions, bounds, special cases, and optimization.
May 2019. Available on arXiv: 1905.13743.

\bibitem{GCSI}
W. Gao, G. Cao, M. Srivatsa, and A. Iyengar. Distributed maintenance of cache freshness in opportunistic mobile networks.
In IEEE ICDCS, June 2012.

\bibitem{YCYW}
R. D. Yates, P. Ciblat, A. Yener, and M. Wigger. Age-optimal constrained cache updating. In IEEE ISIT, June 2017.

\bibitem{KKNWE}
C. Kam, S. Kompella, G. D. Nguyen, J. Wieselthier, and A. Ephremides. Information freshness and popularity in mobile
caching. In IEEE ISIT, June 2017.

\bibitem{ZYS}
J. Zhong, R. D. Yates, and E. Soljanin. Two freshness metrics for local cache refresh. In IEEE ISIT, June 2018.

\bibitem{ZLLGZS}
S. Zhang, J. Li, H. Luo, J. Gao, L. Zhao, and X. S. Shen. Towards fresh and low-latency content delivery in vehicular
networks: An edge caching aspect. In IEEE WCSP, October 2018.

\bibitem{TCWWY}
H. Tang, P. Ciblat, J. Wang, M. Wigger, and R. D. Yates. Age of information aware cache updating with file- and
age-dependent update durations. In International Symposium on Modeling and Optimization in Mobile, Ad Hoc, and Wireless Networks (WiOPT), June 2020.

\bibitem{YZZJ}
L. Yang, Y. Zhong, F. Zheng, and S. Jin. Edge caching with real-time guarantees. December 2019. Available on
arXiv:1912.11847.

\bibitem{BU1}
M. Bastopcu and S. Ulukus. Information freshness in cache updating systems. April 2020. Available on arXiv:2004.09475.

\bibitem{BU2}
M. Bastopcu and S. Ulukus. Information Freshness in Cache Updating Systems with Limited Cache Storage Capacity. May 2020. Available on arXiv:2005.10683.

\bibitem{PAGBAR}
J. Pedersen, A.G. Amat, J. Goseling, F. Brannstrom, I. Andriyanova, and E. Rosnes. Dynamic Coded Caching in Wireless Networks. February 2020. Available on arXiv:2002.08080.

\bibitem{KSUSM}
I. Kadota, A. Sinha, E. Uysal-Biyikoglu, R. Singh, and E. Modiano. Scheduling policies for minimizing age of information
in broadcast wireless networks. IEEE/ACM Transactions on Networking, vol.26, no.6, pp.2637-2650, December 2018.

\bibitem{BU3}
M. Bastopcu and S. Ulukus. Minimizing age of information with soft updates. Journal of Communications and Networks,
vol. 21, no.3, pp.233-243, June 2019.

\bibitem{BSU1}
B. Buyukates, A. Soysal, and S. Ulukus. Age of information scaling in large networks with hierarchical cooperation. In
IEEE Globecom, December 2019.

\bibitem{BU4}
M. Bastopcu and S. Ulukus. Who should Google Scholar update more often? In IEEE Infocom, July 2020.

\bibitem{ZSY}
J. Zhong, E. Soljanin, and R. D. Yates. Status updates through multicast networks. In Allerton Conference, October 2017.

\bibitem{BSU2}
B. Buyukates, A. Soysal, and S. Ulukus. Age of information in two-hop multicast networks. In Asilomar Conference, October 2018.

\bibitem{BSU3}
B. Buyukates, A. Soysal, and S. Ulukus. Age of information in multihop multicast networks. Journal of Communications
and Networks, vol.21, no.3, pp.256-267, July 2019.

\bibitem{ZCV}
C. Zhong, M. Cenk Gursoy, and S. Velipasalar. Deep Multi-Agent Reinforcement Learning Based Cooperative Edge Caching in Wireless Networks. In IEEE ICC, May 2019.

\bibitem{CGG}
E. T. Ceran, D. Gunduz, and A. Gyorgy. A reinforcement learning approach to age of information in multi-user networks.
In IEEE PIMRC, September 2018.

\bibitem{BPLPS}
L. Breslau, P. Cao, L. Fan, G. Phillips, S. Shenker. Web Caching and Zipf-like Distributions: Evidence and Implications. In  Proceedings of IEEE INFOCOM, pp.l126-134, 1999.

\end{thebibliography}
\end{document}